\documentclass[prd,preprint]{revtex4-1}
\usepackage{amsmath,amsfonts,amssymb,hyperref}
\usepackage[english]{babel}
\usepackage[utf8]{inputenc}
\usepackage{amsthm}
\usepackage{balance}
\usepackage{mathtools}
\usepackage{physics}
\usepackage{xcolor}
\usepackage{graphicx}
\usepackage[left=23mm,right=13mm,top=35mm,columnsep=15pt]{geometry} 
\usepackage{adjustbox}
\usepackage{placeins}
\usepackage[T1]{fontenc}
\usepackage{lipsum}
\usepackage{csquotes}
\usepackage{verbatim}

\def\be{\begin{equation}}
\def\ee{\end{equation}}
\def\ber{\begin{eqnarray}}
\def\eer{\end{eqnarray}}
\def\bwt{\begin{widetext}}
\def\ewt{\end{widetext}}

\begin{document}
\title{Anisotropic particle creation from $T-$vacuum in the radiation dominated universe}

\author{Dhamar S. Astilla}    
    \email{dsa1022@usnh.edu}
    \affiliation{University of New Hampshire, 9 Library Way, Durham, NH 03824, USA}

\author{Sujoy K. Modak}    
    \email{smodak@cpp.edu}
     \affiliation{California State Polytechnic University, 3801 West Temple Ave., Pomona, CA 91768, USA}

\author{Enrique Salazar}
    \email{A377773@alumnos.uaslp.mx}
    \affiliation{Instituto de Física, Universidad Aut\'onoma de San Luis Potos\'i, C.P. 78295, San Luis Potos\'i, M\'exico}

\date{\today} 

\begin{abstract}
We further investigate novel features of the $T-$vacuum state, originally defined in the context of quantum field theory in a (1+1) dimensional radiation dominated universe \cite{Modak:2018usa}. Here we extend the previous work to a realistic (3+1) dimensional set up and show that $T-$vacuum gives rise to an \emph{anisotropic particle creation phenomena} in the radiation dominated early universe. Unlike the Hawking or Unruh effect, where the particle content is thermal and asymptotically defined, here it is {\it non-thermal and instantaneous}. This novel example of particle creation is interesting because these particles are detected in the frame of physical/cosmological observers, who envision $T-$vacuum as a particle excited state. Such results comes with a potential to be eventually compared with the observed anisotropies from the early universe and may provide new insights on cosmological particle creation.

\end{abstract}


\maketitle

\tableofcontents

\section{Introduction}
Within the framework of Quantum Field Theory in Curved Spacetime (QFTCS), under certain symmetries of the spacetime, there may exist non-unique, unitarily in-equivalent decomposition of quantum fields, which are usually expressed using distinct coordinates. It is not guaranteed that every curved spacetime would offer such quantizations in Fock space, but when they do,  a direct notable consequence of this is the existence of non-unique vacuum states corresponding to different observers who are, in general, non-inertial with respect to each other. Therefore, the meaning of \textit{particle} is not universally accepted -- vacuum states for a set of observers could be manifested as particle excited quantum states for other set of observers. The only observers that agree with the notion of particle are the \textit{inertial observers} in the Minkowski/flat spacetime, who are related by Lorentz transformations. However, no such unique global notion of particles can be attributed for observers in a curved spacetime including the cosmological setting.

This novel difference between the standard description of quantum field theory (QFT) in a flat and curved space has inspired a generation of physicists to look into various aspects of QFT in curved space and obtaining interesting results of particle excitation in various settings. Among them, most well known are particle creation by black holes \cite{hawk1, hawk2}, in a  cosmological settings \cite{park1}-\cite{park3} as well as for the accelerated observers in flat spacetime (i.e., the Unruh effect \cite{unruh}). The work of Unruh pointed out that an accelerated observer in Minkowski spacetime (with uniform four-acceleration $a^{\mu}$) detects particles which are created by the excitation of the Minkowski vacuum. Further, the resulting radiation flux follow a thermal distribution with a well-defined temperature $T=|a|/2\pi$ (in natural units). It is understood that the energy required to put the observer in an accelerated motion is also responsible for disturbing the Minkowski vacuum and produce particles. Although important, the details of how to put the observer and the detector in accelerated motion is not essential for understanding the basic result of the Unruh effect. A nice review of Unruh effect and related works can be found in \cite{rev2}.

Recently, one of us showed \cite{Modak:2018usa} that an analogous situation can be contemplated in the radiation dominated early universe, where physical/cosmological observers detect radiation flux  as a result of an excitation of a novel ``$T-$vacuum state''. This is analogous to the case where accelerated observers detect particles created by the Minkowski vacuum state. Although, similar, there exists an important difference with the Unruh effect -- in Unruh effect the source of energy driving the accelerated observer is not included, but in the radiation dominated universe there is a gravitational force and various observers have their own proper frames that are not globally inertial. In fact, this $T-$vacuum is a natural state for a new set of observers first introduced in \cite{Modak:2018crw} who experience the radiation dominated universe as a spherically symmetric metric, inhomogeneous and anisotropic spacetime. Various geometric aspects of the said metric were also highlighted in \cite{Modak:2018crw, Modak:2018usa}. In \cite{Salazar:2021pfm}, in a relatively different study, we provided a cosmic identity of the observers for whom such a metric appear natural. We showed that the natural observers inhabiting such a spherically symmetric form of the metric can be backtracked and identified with the static de Sitter observers in the previous (inflationary) epoch. In other words, $T-$vacuum state turns out to be also a natural vacuum state for the static de Sitter observers when, the universe has made a transition to the next (radiation) epoch and, the spacetime is no longer inflationary but dominated by radiation. On the field theory side, regarding particle creation, a point-wise comparison was also made in a (1+1) dimensional case \cite{Modak:2018usa} to clearly state the similarity and differences with the Unruh effect. These studies on the geometric and field theoretic aspects of the radiation dominated universe is  a part of a broader motivation of understanding various subtleties associated with the QFT in a realistic universe with multiple epochs \cite{Singh:2013bsa, Modak:2019jbg} of expansion. In \cite{Modak:2018usa} we also showed that $T-$vacuum state is Hadamard (no divergences or other pathological behaviors present) in (1+1) dimensions. 

While the geometric construction of the above studies in the context of the radiation dominated universe was performed considering (3+1) dimensions, the field theoretic formulation on particle creation and renormalization was confined only in (1+1) dimensions. In this work we extend the particle creation by $T-$vacuum in a (3+1) dimensional set-up and leave the renormalization properties for a future work \cite{ams}. First, considering a massless scalar field we derive various field equations and quantize the field separately in the cosmological and spherically symmetric metrics describing the radiation dominated universe. Some relational aspects of the Hubble horizon in two metrics are carefully mentioned. We then explicitly calculate the Bogolyubov coefficients and particle number density within the $s-$wave approximation and compare the result with the two dimensional case. Finally, we consider waves with arbitrary $l> 0$ and manage to calculate particle number densities {\it analytically} for the deep sub-Hubble modes. Naturally, the particle distribution is a function of the conformal time $\eta$ (or the scale factor $a$). Our result thus confirm an anisotropic, time-dependent particle creation by $T-vacuum$ in the early universe and in the radiation dominated universe. In principle, this particle creation may have a imprint  on the Cosmic-Microwave-Background (CMB) especially because it  takes place after the inflation and there is no such process (like inflation)  which may completely wipe out residuals of these tiny anisotropies. It will also indeed be interesting to see what other  cosmological implications may arise due to this gravitational particle creation. In this work we do not delve into such issues,  rather set a solid platform where further questions may be asked and we shall address them separately elsewhere \cite{ams}. 

The paper is organized as follows: in section \ref{sec2} we present a brief review of various geometric features related to the spherically symmetric form of the radiation dominated universe. It includes the coordinate transformation, identification of cosmological observers' trajectory, spacetime foliation, comparison of Hubble radius and connection with static de Sitter metric. In the following sections, \ref{sec-3} and \ref{sec-4}, we quantize massless scalar field in the two frames relevant for this study. Section \ref{sec-5} provides the main result of particle creation separately for the $s-$wave approximation and beyond the $s-$wave. Finally, we conclude in \ref{sec-6}. There are two useful appendices \ref{ap-a} and \ref{ap-b} where we calculate the normalization constants and Bogolyubov coefficients.


\section{Novel geometric features of the radiation dominated early universe}\label{sec2}

In this section we review the key results of some of the our recent works which are essential for the main study of particle creation in the radiation dominated universe to be presented in the subsequent sections of this article.

\subsection{Conformal vs spherically symmetric forms of the radiation stage}

Let us start from the spatially flat FRW metric in comoving coordinates, 
\begin{equation}\label{1}
    ds^2 = dt^2 - a^2(t)[dr^2 + r^2(d\theta^2 + \sin^2\theta d\phi^2)],
\end{equation}
which can be expressed  in a conformal flat form 
\begin{equation}\label{2}
    ds^2 = a^2(\eta)[d\eta^2 - dr^2 + r^2(d\theta^2 + \sin^2\theta d\phi^2)].
\end{equation}
which we call here the ``cosmological frame'' $(\eta, r, \theta, \phi)$ where $\eta = \int \frac{dt}{a(t)}$ is the conformal/cosmological time. In our convention the scale factor $a(t)$ has a dimension of length while $\eta$ is dimensionless ($t$ has dimension of time). Thus except the conformal factor  everything is dimensionless in \eqref{2}.

We are working in a radiation dominated stage of the universe which comes after the inflationary universe. Considering scale factors for these two stages one can determine a time where this transition takes place \cite{Singh:2013bsa}. The idea is to make the scale factor equal and differentiable at some transition point. that is, if we consider $a_{inf}=e^{{\cal H}t}$ and $a_{rd}=a_{0} t^{1/2}$ and make them continuous and differentiable at some $t=t_r$, we can fix $t_r=1/2{\cal H}$ and  $a_{0} = \sqrt{2\mathcal{H}e}$ (where $\mathcal{H}$ is the inflationary Hubble constant). Substituting both we can obtain the value of scale factor $a_{inf}=a_{rd}=\sqrt{e}$, the length scale where the early universe transits from the inflationary to the radiation dominated stage. The metric \eqref{2} will later be used for the construction of a quantum field theory. 

Now we want to express the radiation dominated universe in a spherically symmetric and conformally static form which was discovered in \cite{Modak:2018crw}. First rewrite \eqref{2} by defining the lightcone coordinates $u = \eta - r$ and $v = \eta + r$, so that equation (\ref{2}) becomes
\begin{equation}
    ds^2 = a^2(u,v)\Big[dudv - \frac{(v-u)^2}{4}(d\theta^2 + \sin^2\theta d\phi^2)\Big].
\end{equation}
Then perform the following conformal transformation of the cosmological null coordinates for $a(t) \propto t^{1/2}$, given by \cite{Modak:2018crw}
\begin{equation}
    U \equiv T-R = \pm \frac{\mathcal{H}e}{2}u^2, \ \ \ V \equiv T+R = \frac{\mathcal{H}e}{2}v^2,
\end{equation}
where $+$ and $-$ signs stand for $u\geq0$ and $u\leq0$, respectively. This transformation gives the following metrics for the sub-Hubble ($I$) and super-Hubble ($II$) regions, described by
\begin{equation}
    ds^2 = F_I(T, R)(dT^2 - dR^2) - R^2d\Omega^2
    \label{stsub}
\end{equation}
for $U \geq 0$ ($T\geq R$), and
\begin{equation}
    ds^2 = F_{II}(T,R)(dT^2 - dR^2) - T^2d\Omega^2, 
    \label{stsup}
\end{equation}
for $U\leq 0$ ($T\leq R$). With the functions $F_I$ and $F_{II}$ as
\begin{equation}
    F_{I}(T,R) = \frac{(\sqrt{T+R} + \sqrt{T-R})^2}{4\sqrt{T^2-R^2}}
\end{equation}
\begin{equation}
    F_{II}(T,R) = \frac{(\sqrt{R+T} - \sqrt{R-T})^2}{4\sqrt{R^2-T^2}}
\end{equation}
The relationship between $(T,R)$ and $(\eta,r)$ frames is given by 
\begin{equation}\label{TransformationsRegion1}
    \begin{aligned}
    T &= \frac{V+U}{2} = \frac{\mathcal{H}e}{2}(\eta^2 + r^2) \\
    R &= \frac{V-U}{2} = \mathcal{H}e\eta r
    \end{aligned} 
\end{equation}
\\
for region $I$, and 
\begin{equation}\label{TransformationRegion2}
    \begin{aligned}
    T &= \frac{V+U}{2} = \mathcal{H}e\eta r \\
    R &= \frac{V-U}{2} = \frac{\mathcal{H}e}{2}(\eta^2 + r^2)
    \end{aligned} 
\end{equation}
for region $II$.

We can also express the conformal factors $F_{I}(T,R)$ and $F_{II}(T,R)$ as functions of the Hubble parameter for radiation stage $H = (\frac{\Dot{a}}{a})_{RD}$ \cite{Modak:2018crw}. These are
\begin{equation}
    F_{I}(H,R) = \frac{1}{1-H^2R^2}
\end{equation}
and
\begin{equation}
    F_{II}(H,T) = \frac{1}{H^2T^2-1}.
\end{equation}
\\
The lightcone boundary $T = R$ for these observers is translated as the comoving Hubble radius at  $R=1/H$.

These new metrics \eqref{stsub} and \eqref{stsup} are static up to a dynamical conformal factor and they exhibit a spherical symmetry. We shall refer to this coordinate system as the $(T,R)$ frame to meaningfully express our thoughts. The spherical symmetry in $(T,R)$ frame have some interesting geometric and field theoretic implications. We shall discuss them in detail in the remaining part of this work.

\subsection{Identification of cosmological observers in the $(T,R)$ frame}
\label{sec3}
We want to identify the physical cosmological observers in the $(T,R)$ since their frame is often used in cosmology. It was shown in \cite{Modak:2018usa} that for the sub-Hubble region ($I$) a worldline of the form 
\begin{equation}
    T = G(R) = \alpha_1 R^2 + \beta_1
\end{equation}
with $\beta_1 = 1/4\alpha_1 = \mathcal{H}e\eta^2_0/2$ is identified as the cosmological static frame (at rest at an arbitrary but sub-Hubble point $r=r_0$). 
These observers have a radial velocity $\frac{dR}{dT} = \frac{1}{2\alpha_1R}$ and a non-zero position dependent deceleration $\frac{d^2R}{dT^2}= -\frac{1}{4\alpha^2_1R^3}$.   
For the super-Hubble region ($II$) cosmological observers at constant $r = r_0$ are described by the following worldline
\begin{equation}
    R = G(T) = \alpha_1T^2 + \beta_1
\end{equation}
with $\beta_1 = 1/4\alpha_1 = \mathcal{H}er^2_0/2$. The radial velocity is time dependent, $\frac{dR}{dT} = 2\alpha_1T$, while the acceleration is constant $\frac{d^2R}{dT^2} = 2\alpha_1$ for this case.
Therefore, if we see the journey of the cosmological observer, in $(T,R)$ frame, starting from  the super-Hubble region, we find the said observers are accelerating at a constant rate which lasts up to the Hubble entry. After the Hubble entry they start decelerating and  come to a rest, even in $(T,R)$ frame, as they approach $\mathcal{I}^+$. As expected, the cosmological observers do not encounter any horizon (due to a coordinate singularity) although the metric \eqref{stsub} and \eqref{stsup} is fragmented into the sub-Hubble and super-Hubble regions.

\subsection{Cosmological foliation in $(T,R)$ frame}\label{sec4}

While cosmological frame in $(\eta, r, \theta, \phi)$ has a straightforward foliation of time and space slices with $\eta=$const. and $r=$const., one has a slightly nontrivial case in $(T,R)$ frame where new time and space slices are identified with $T=$const. and $R=$const. For a well defined initial value formulation it is necessary that any spacetime is foliated by Cauchy slices. Since we already have Cauchy problem well defined in the cosmological frame we can prove that the same is true in the $(T,R)$ frame if we can embed cosmological time and space slices using the $(T,R)$ metric. This task was also completed in \cite{Modak:2018usa}.

The first step is to identify timeslices $\eta = \eta_0$, for the sub-Hubble region, 
\begin{equation}\label{2.3}
    T = \frac{R^2}{2\mathcal{H}e\eta^2_0} + \frac{1}{2}\mathcal{H}e\eta^2_0,
\end{equation}
\noindent 
taking into account only the portion below the Semi-Latus-Rectum (SLR) of the above parabola. For the super-Hubble region the timeslices will be a portion of the following parabola 
\begin{equation}
    R = \frac{T^2}{2\mathcal{H}e\eta^2_0} + \frac{1}{2}\mathcal{H}e\eta^2_0
\end{equation}
starting from the point where the SLR of (\ref{2.3}) meet the Hubble horizon. Joining these two slices along the Hubble horizon provides an embedding of the full cosmological time slice in the $(T,R)$ frame. The spaceslices (timelike hypersurfaces) are constructed in a similar way, just by taking $r = r_0$. In region-$I$ the spaceslice is given by
\begin{equation}\label{2.5}
     T = \frac{R^2}{2\mathcal{H}er^2_0} + \frac{1}{2}\mathcal{H}er^2_0
\end{equation}
\noindent
while for the region-$II$ the spaceslice is
\begin{equation}\label{2.6}
    R = \frac{T^2}{2\mathcal{H}er^2_0} + \frac{1}{2}\mathcal{H}er^2_0,
\end{equation}
again joining at the Hubble radius. The union of these two portions is made by joining the upper portion (of the SLR) of the parabola (\ref{2.5}) with the left portion (of the SLR) of the parabola (\ref{2.6}). 

\begin{figure}[t]
    \centering
    \includegraphics[width=0.7\linewidth]{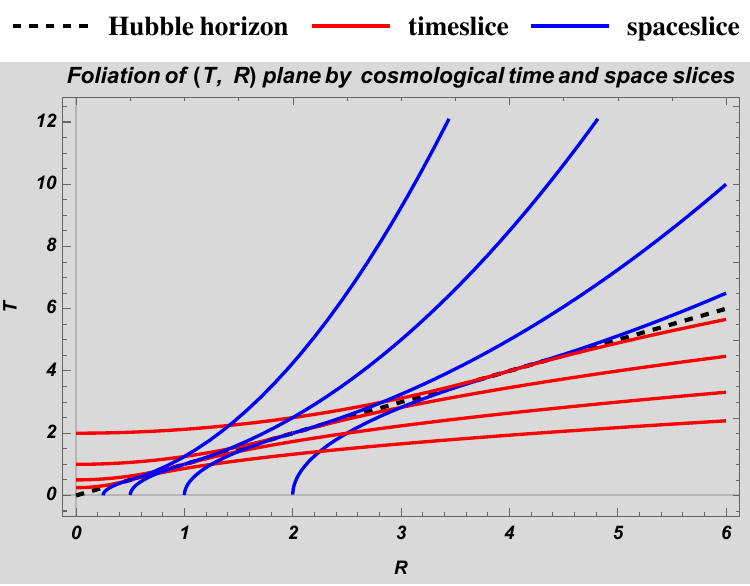}
    \caption{Full spacetime foliation of the $(T,R)$ plane with respect to cosmological time and spaceslices as shown in \cite{Modak:2018usa}. The blue slices are the geodesics for a cosmological observer with $r=r_0=\text{const.}$ and the red ones are the spacelike trajectories ($\eta = \eta_0 = \text{const.}$). The dashed line is the comoving Hubble radius.}
    \label{FullSpacetimeFoliation}
\end{figure}

\ref{FullSpacetimeFoliation}. 

With these cosmological Cauchy slices we can now foliate the full spacetime in the $(T,R)$ plane. There is no horizon for cosmological observers as expected. Looking into the blue curves in the diagram \ref{FullSpacetimeFoliation}, now we can confirm that cosmological observers start radially accelerating in asymptotic past while looked at from a $(T,R)$ frame which is valid in the super-Hubble region. After entering the Hubble radius they start decelerating and come to a rest in the asymptotic future at $\mathcal{I}^+$. Here we first get an impression that a vacuum state in $(T,R)$ frame cannot be vacuum for cosmological observers since these observers are non-inertially connected with each other.

\subsection{Hubble parameter and horizon behaviour in $(T,R)$ frame}
\label{sec5}

It is important to realize that the transformed metrics \eqref{stsub} and \eqref{stsup} representing the radiation dominated universe are both inhomogeneous and anisotropic since they dependent on both $T$ and $R$. A direct consequence of that will be the value of the Hubble parameter depending not only on new time $T$ but also on the position of the observer $R$. We expect the Hubble horizon to be manifested in a nontrivial manner while using these new coordinates.

Hubble parameter is calculated from the knowledge of the scalar factor $a(t)$ and its derivative ($H = \frac{\Dot{a}}{a}$), meaning that an observer in any part of the universe can measure the same value if they are in the same timeslice. In the case of the radiation dominated universe we have $H(t) = \frac{1}{2t}$, but since we are working on the cosmological frame ($\eta = \frac{2}{a_0}t^{1/2}$, where we are using $a_0=\sqrt{2\mathcal{H}e}$) then the value of this parameter becomes
\begin{equation}\label{HubbleEta}
    H = \frac{2}{a^2_0\eta^2} = \frac{1}{\mathcal{H}e\eta^2}.
\end{equation}

Since $\eta$ has different expressions in sub and super Hubble regions due to (\ref{TransformationsRegion1}) and (\ref{TransformationRegion2}), in the $(T,R)$ frame we will have different expressions for the Hubble parameter in those regions. In region $I$, using the coordinate transformations in (\ref{TransformationsRegion1}), we can express $\eta$ as
\begin{equation}
    \eta = \frac{1}{\sqrt{2\mathcal{H}e}}\Big(\sqrt{T+R}+\sqrt{T-R}\Big).
\end{equation}
Using (\ref{HubbleEta}) and simplifying we obtain 
\begin{equation}
    H_{I}(T,R) = \frac{1}{T+\sqrt{T^2-R^2}}
\end{equation}
for the region $I$, i.e., for $R<T$. For the region $II$, we use coordinate transformation given in (\ref{TransformationRegion2}), and obtain a similar expression for $\eta$, and by doing a similar process, we can obtain the value of the Hubble parameter for this region, as
\begin{equation}
    H_{II}(T,R) = \frac{1}{R + \sqrt{R^2-T^2}}.
\end{equation}
This expression is valid for $R>T$. 

We can easily observe that these expressions are continuous on $T = R$, coinciding in the value $H = \frac{1}{T}= \frac{1}{R}$. Also we can observe, as we expected, that these expressions are function of $T$ and $R$ coordinates, so in general any two observers in different points on space or time measure different values of Hubble parameter, this is due to the nature of the metric which defines the $(T,R)$ spacetime which breaks homogeneity and isotropy.



As we have the expression for the Hubble parameter in terms of new $T$ and $R$ coordinates, we can express the radii for these two regions in both the sub-Hubble and super-Hubble spacetime sections which are simply
\begin{equation}
    L^{I}_H = L^I_P = T_0+\sqrt{T^2_0-R^2_0}
\end{equation}
for region-I ($R<T$) and
\begin{equation}
    L^{II}_H = L^{II}_P = R_0+\sqrt{R^2_0-T^2_0}
\end{equation}
for the region-II ($R>T$). We can observe that now we need to measure both the current time and radial position to get radii values instead of just measuring the current time in cosmological frame. In the new frame the Hubble radius and particle horizon depend on the position where and when they are measured.

\subsection{The relationship of $(T,R)$ frame with the static de Sitter frame}

In a previous work involving one of us \cite{Salazar:2021pfm} we showed an interesting property of the metric \eqref{stsub} representing the $(T,R)$ frame. The point is \eqref{stsub} can be backtracked and matched with the static de Sitter frame in the preceding inflationary epoch with a time redefinition. Therefore, static de Sitter observers in the inflationary epoch will find themselves closely related with the $(T,R)$ frame when a transition to the radiation epoch has already taken place. This identification will be helpful in a way that will be made more clearer later on when we discuss particle creation. Most importantly, this observation provides a valuable ``cosmic identity'' of these new observers in the radiation dominated universe. To demonstrate how this happens mathematically it was shown in \cite{Salazar:2021pfm} that de Sitter universe in the static frame, given by
\begin{equation} \label{st-ds}
   ds^2 = (1 - {\cal H}^2 R^2) {d{\tau}^2} - \frac{dR^2}{1 - {\cal H}^2 R^2} - R^2 d\Omega^2, 
\end{equation}
smoothly transits into the following metric in the radiation epoch,
\begin{equation} \label{st-dsr}
 ds^2 =  \frac{4 e^2 e^{4{\cal H} \tau}d\tau^2 - dR^2}{1 - H^2 R^2} - R^2 d\Omega^2.
\end{equation}
a transition which takes place at the comoving time $t_r=1/2{\cal H}$ which in $\tau,R$ coordinates is translated to $R=a(t_r)r$ and $\tau = \frac{1}{2{\cal H}}\log|\frac{1}{2}({\cal H}^2r^2-1/e)|$. We may call \eqref{st-dsr} as the de Sitter ``post-static'' frame as described in the radiation stage of the universe. We suggest an interested reader to read section 5 of \cite{Salazar:2021pfm} for technical details. The static de sitter metric, as represented by \eqref{st-dsr}, has a striking similarity with the metric \eqref{stsub}. If we redefine the de Sitter time 
\begin{equation}\label{redf}
\tau = \frac{1}{2{\cal H}} \log({{\cal H}T}/{e})
\end{equation}
then we see that \eqref{st-dsr} matches exactly with \eqref{stsub}.  Therefore, one may interpret that \eqref{stsub} is actually the de Sitter static frame living in the radiation epoch with a logarithmic scaling of time. Note that \eqref{redf} is purely temporal and does not mix the time with space coordinates which will have a crucial role to interpret the particle creation in even in this de Sitter static post-static frame, to be discussed in the subsequent sections of the paper.


\section{Field modes in conformally flat FRW}
\label{sec-3}

Now that we have a geometric understanding of the radiation dominated universe in both frames we turn our attention on the construction of quantum field theory on them. In this section we consider the cosmological frame \eqref{2} where the metric is conformally flat.

 Consider a massless scalar field with arbitrary coupling in the radiation stage and solve the Klein-Gordon equation $\Box \Phi = 0$ (simply because $R=0$ and all fields become minimally coupled). We can express the scalar field ansatz using the separation of variables as 
\begin{equation}
    \Phi(\eta,r,\theta,\phi) = \sum_{\ell,m} \frac{f_\ell(r)}{r}g(\eta)Y_{\ell m}(\theta,\phi)
\end{equation}
\noindent
where the angular part $Y_{\ell m}$ are the spherical harmonics and the $(\eta,r)$ dependent parts satisfy the following equations 
\begin{equation}\label{3.2}
    \eta^2\frac{d^2g}{d\eta^2} + 2\eta\frac{dg}{d\eta} + \omega^2\eta^2g = 0,
\end{equation}
\begin{equation}\label{3.3}
\frac{d^2f_\ell}{dr^2} + \Bigg(\omega^2 - \frac{\ell(\ell+1)}{r^2}\Bigg)f_\ell = 0.
\end{equation}
\noindent
The time dependent variable $g(\eta)$ satisfies the well-known spherical Bessel equation and we can use the Fourier mode decomposition and then sum over all modes to calculate $g(\eta)$. 

At this point, we shall separate our study for the $\ell=0$ or the $s-$waves and for higher values of $\ell>0$ to make the comparison clear with the spherically symmetric case.

\subsection{Field modes for the $s-$wave approximation}
Considering the s-wave approximation ($\ell=0$) in (\ref{3.3}) and solving the differential equations we can express the field $\Phi$ as 

\begin{equation}
    \begin{aligned}
    \Phi &= \sum_\omega \mathcal{N}\frac{f_0(r)}{r}g_\omega(\eta)Y_{00}(\theta,\phi)  \\
    & = \sum_\omega \mathcal{N} \Bigg(\frac{e^{-i\omega(\eta-r)}}{2\sqrt{\pi}\omega\eta r}a^{in}_\omega + \frac{e^{-i\omega(\eta+r)}}{2\sqrt{\pi}\omega\eta r}a^{out}_\omega\Bigg) + \text{h.c.} \label{SmodeFlatFLWR}
    \end{aligned}
\end{equation}
where $\mathcal{N}$ is a normalization constant determined by the orthogonality condition between the field modes which is simply $\mathcal{N} = \frac{\sqrt{\omega}}{\sqrt{4\pi}\mathcal{H}e}$. The superscripts {\it in} and {\it out} implies the usual ingoing and outgoing modes. The field expansion for $\Phi$ in the integral form can be written as
\begin{equation}\label{FieldExpansionCosmological}
    \Phi = \int d^3\omega (u_\omega a^{in}_\omega + v_\omega a^{out}_\omega) + \text{h.c.}
\end{equation}
with the field modes $u_\omega$ and $v_\omega$ being
\begin{equation}\label{uMode}
    u_\omega = \frac{e^{-i\omega(\eta-r)}}{4\pi \mathcal{H}e\sqrt{\omega}\eta r}
\end{equation}
and
\begin{equation}\label{vMode}
    v_\omega = \frac{e^{-i\omega(\eta+r)}}{4\pi \mathcal{H}e\sqrt{\omega}\eta r}.
\end{equation}

\subsection{Field modes beyond the $s-$wave approximation}\label{sec-3b}

The general solution for equation (\ref{3.3}) when $\ell \neq 0$ has the form of 
\begin{equation}
    f(r) = \sqrt{r}(C_1 J_{\ell+1/2}(\omega r) + C_2 Y_{\ell +1/2}(\omega r) )
\end{equation}
where $J_{\ell +1/2}$ and $Y_{\ell + 1/2}$ are Bessel functions of first and second kind, respectively. The Bessel function of second kind $Y_{\ell+1/2}(r)$ blows up at $r\to 0$, so we must choose $C_2 = 0$ and let us rename $C_1={\mathcal{N}}$ so that that 
\begin{equation}
    f_\ell(r) = \mathcal{N}\sqrt{r} J_{\ell +1/2}(\omega r),
\end{equation} 
where $\mathcal{N}$ is the normalization constant determined by the orthogonality condition between the field modes. The value of $\mathcal{N}$ is calculated in Appendix \ref{ap-a}, giving $\mathcal{N} = {\omega }/{\sqrt{2} \mathcal{H}e}$.

For further simplification, the Bessel function can be separated in terms in spherical Hankel functions such that 
\begin{align*}
    J_{\ell + 1/2}(\omega r) &= \sqrt{\frac{2\omega r}{\pi}} j_\ell(\omega r)  \\
    & = \sqrt{\frac{2\omega r}{\pi}} \frac{1}{2} \left( h^{(1)}_\ell(\omega r) + h^{(2)}_\ell(\omega r) \right) \\
    & = \sqrt{\frac{\omega r}{2 \pi}} \left(\frac{e^{i \omega r}}{\omega r} Q_{\ell}(\omega r) + \frac{e^{-i \omega r}}{\omega r} Q_{\ell}^*(\omega r) \right)
\end{align*}
Where the spherical Hankel functions were expanded as
\begin{align}\label{qell}
	h^{(1)}_\ell (x) &=(-i)^{\ell+1} \frac{e^{i x}}{x} \sum_{s=0}^{\ell} \frac{(i/2)^s}{s!} \frac{(\ell+s)!}{(\ell-s)!} x^{-s} \equiv \frac{e^{i x}}{x} Q_{\ell}(x),\\
	h^{(2)}_\ell (x) &= (i)^{\ell+1} \frac{e^{-i x}}{x} \sum_{s=0}^{\ell} \frac{(-i/2)^s}{s!} \frac{(\ell+s)!}{(\ell-s)!} x^{-s} = \frac{e^{-i x}}{x} Q_{\ell}^*(x)
\end{align}
Then, the complete field expansion is given by
\begin{widetext}
\begin{equation}\label{fideo}
    \Phi =  \frac{1}{\mathcal{H}e \eta} \sum_{\ell,m} \int d^3\omega \,  \sqrt{\frac{\omega}{4\pi}} \left(a^{in}_{\omega l m} \frac{e^{-i\omega(\eta - r)}}{\omega r} Q_{\ell}(\omega r) + a^{out}_{\omega l m} \frac{e^{-i\omega(\eta + r)}}{\omega r} Q_{\ell}^*(\omega r) \right) Y_{\ell m}(\theta,\phi) + h.c.
\end{equation}
\end{widetext}
From the above expression we can read-off the mode functions
\begin{eqnarray}
  u_{\omega lm} &=&  \sqrt{\frac{\omega}{4\pi}} \frac{e^{-i\omega u}}{\mathcal{H}e\omega \eta r} Q_{\ell} (\omega r) Y_{\ell m}(\theta,\phi), \\
  v_{\omega lm} &=&  \sqrt{\frac{\omega}{4\pi}} \frac{e^{-i\omega v}}{\mathcal{H}e\omega \eta r} Q^*_{\ell} (\omega r) Y_{\ell m} (\theta,\phi).
\end{eqnarray}
The above two equations show that unlike the $s-$wave sector, here the complete set of modes cannot  be separated into a incoming and outgoing  sectors. The functions $Q_{\ell} (\omega r)$ and $Q_{\ell}^*(\omega r)$ are multiplying the incoming and outgoing modes which is expected for a situation where there is no spherical symmetry. This process of back-scattering is nicely captured by $Q_{\ell}/Q_{\ell}^*$ multiplication to the incoming and outgoing sectors.

\section{Field modes in spherically symmetric FRW}
\label{sec-4}
In this section we quantize massless scalar fields in the background geometries, given by \eqref{stsub} and \eqref{stsup} representing spherically symmetric forms of the radiation dominated universe in the sub and super Hubble scales. We present this calculation separately using the above metrics as backgrounds, also separating the $s-$waves and $\ell>0$ cases.

\subsection{Sub-Hubble region}
Since \eqref{stsub} is spherically symmetric, we can write the scalar field ansatz as
\begin{equation}
    \Phi_\Omega = \sum_{\ell,m}\frac{\Phi^{\ell m}_\Omega(T,R)}{R}Y_{\ell m}(\theta, \phi)
\end{equation} 
Then, the Klein-Gordon equation is
\begin{equation}
    \Bigg(\frac{\partial^2\Phi^{\ell m}_\Omega}{\partial T^2}- \frac{\partial^2\Phi^{\ell m}_\Omega}{\partial R^2}\Bigg) + \frac{\ell(\ell+1)}{R^2(1-H^2R^2)}\Phi^{\ell m}_\Omega = 0,
    \label{sh}
\end{equation}
where we find an effective potential
\begin{equation}
    V_{\ell} (R,H) = \frac{\ell(\ell+1)}{R^2(1-H^2R^2)}.
    \label{pot}
\end{equation}
For any $\ell\neq 0$ modes, this potential has an interesting behavior -- it diverges at $R=0$ and $R=1/H$, i.e., at the origin of the coordinate system and also at the Hubble radius. While the divergence at the origin is a characteristic of the spherical waves because of the choice of coordinate system, the other divergence at $R=1/H$ is nontrivial. We need to understand this in a finer detail. We know that whenever the potential tends to infinity at some value of the coordinate, it means that the mode functions are confined within this limit. That is, no $\ell \ne 0$ modes can pass to the super-Hubble region, they will always be trapped inside the Hubble radius, i.e., in the sub-Hubble region. Note that although our background spacetime is defined in the sub-Hubble region, mode functions are global and there is, in principle, no restriction to extend them beyond the Hubble scale. Peculiarity of the potential \eqref{pot} is interesting in this regard. It must also be added that  although in this new coordinates, it may appear that the Hubble horizon is behaving just like a black hole event horizon where modes cannot get outside from the inside of the event horizon, there is a crucial difference. The potential barrier is mode specific -- i.e.,  \eqref{pot} tells us $V_{\ell}(R,H) = 0$  for $\ell=0$ modes and hence the $s-$wave modes see no potential whatsoever but other modes do feel the potential barrier which is not a characteristic of an event horizon.

Given the peculiarity of the potential \eqref{pot} we now part ways in discussing the $\ell=0$ and $\ell>0$ modes in the following discussion.

\subsubsection{Field modes for the s-wave approximation}
If we consider  the $\ell = 0$ modes only we have from \eqref{sh},
\begin{equation}
    \Bigg(\frac{\partial^2\Phi^{00}_\Omega}{\partial T^2}- \frac{\partial^2\Phi^{00}_\Omega}{\partial R^2}\Bigg) = 0.
\end{equation}
The field expansion is of the following form
\begin{equation}\label{FieldExpansionTRFrame}
    \Phi = \int d^3\Omega (U^{\text{sub}}_\Omega b^{in}_{\Omega>\Omega_H}+V^{\text{sub}}_\Omega b^{out}_{\Omega>\Omega_H}) + \text{h.c.}
\end{equation}
where the mode functions are 
\begin{equation}\label{sub-su}
    U^{\text{sub}}_\Omega = \frac{1}{4\pi\sqrt{\Omega}R}e^{-i\Omega(T-R)}
\end{equation}
and 
\begin{equation}\label{sub-sv}
    V^{\text{sub}}_\Omega = \frac{1}{4\pi\sqrt{\Omega}R}e^{-i\Omega(T+R)}
\end{equation}
and are valid for $R\leq 1/H$. The operator $b^{in}_{\Omega>\Omega_H}$ is defined for the incoming sub-Hubble modes and the state which is annihilated by this operator is called the $T-$vacuum (for this sector) which satisfy $b^{in}_{\Omega>\Omega_H}|0 \rangle_{T}^{in} = 0$. Similarly, one can define a vacuum  $b^{out}_{\Omega>\Omega_H}|0 \rangle_{T}^{out} = 0$ for the outgoing sector.

Our focus in this paper will be to calculate particle creation from the $T-$vacuum state when looked into from the cosmological frame described in the last section.

\subsubsection{Field modes beyond the s-wave approximation}
If we want to go beyond the $s-$waves, we need to solve \eqref{sh} for arbitrary $\ell>0$ which is clearly a difficult task since the Hubble parameter depends on space and time $H\equiv H(T,R)$. It makes the differential equation quite nontrivial to solve in an exact form. 


However, we can consider a curious case by restricting our interest to the deep sub-Hubble region $R<<1/H$. In this case \eqref{sh} becomes
\begin{equation}
     \Bigg(\frac{\partial^2\Phi^{\ell m}_\Omega}{\partial T^2}- \frac{\partial^2\Phi^{\ell m}_\Omega}{\partial R^2}\Bigg) + \frac{\ell(\ell+1)}{R^2}\Phi^{lm}_\Omega = 0,
    \label{dsh}
\end{equation}
By decomposing $\Phi^{\ell m}_\Omega(T,R) = \psi(T)\chi(R)$ and solve \eqref{dsh} it is found that 
\begin{align}
    \frac{d^2\psi}{dT^2} & +\Omega^2\psi = 0 \\
    \frac{d^2\chi}{dR^2} & + (\Omega^2 - \frac{\ell(\ell+1)}{R^2})\chi = 0 
\end{align}
it is clear that the solutions for $\psi(T) \propto e^{\pm i \Omega T}$, while the differential equation for $\chi(T)$ it is the exact same form of \eqref{3.3}, and its solution is the Bessel function of first kind $\sqrt{R} J_{\ell + 1/2}(\Omega R)$.

The exact calculation for the normalization constant is now a bit nontrivial. This is because since our field modes are only valid for a restricted region $R<<1/H$ we cannot use them to normalize over the entire spatial volume element within the Hubble radius. Therefore, the normalization constant calculated in the appendix \ref{ap-a}, using above modes are only an approximation. To account for this approximation, we include a factor ${\cal N}_{\text{sub}}$ in the field decomposition, which is now accurately expressed as
\begin{widetext}
\begin{equation}
\label{fide}
    \Phi = \sum_{\ell,m} \int d^3\Omega {\cal N}_{\text{sub}} \,  \sqrt{\frac{\Omega}{4\pi}} \left(b^{in}_{\Omega \ell m} \frac{e^{-i\Omega(T - R)}}{\Omega R} Q_{\ell}(\Omega R) + b^{out}_{\Omega \ell m} \frac{e^{-i\Omega(T + R)}}{\Omega R} Q_{\ell}^*(\Omega R) \right) Y_{\ell m}(\theta,\phi) + \text{h.c.}
\end{equation}
\end{widetext}
Determination of the normalization constant ${\cal N}_{sub}$ would require an integration over the three dimensional volume element and exact mode solutions for all values of $R$. For this work, we stay within the analytical limit and work with unnormalized modes which does not limit our ability to discuss the anisotropic particle creation phenomena from the $T$ vacuum which is our main goal in this work. We plan to solve equation \eqref{sh} numerically to calculate  in a separate work elsewhere. It is sufficient for now to consider the above mentioned  unnormalized mode functions:
\begin{eqnarray}\label{sub-lu}
    U_{\Omega \ell m} &=& \sqrt{\frac{\Omega}{4\pi}}  \frac{e^{-i\Omega U}}{\Omega R} Q_{\ell}(\Omega R) Y_{\ell m}(\theta,\phi) \\
    V_{\Omega \ell m} &=& \sqrt{\frac{\Omega}{4\pi}}  \frac{e^{-i\Omega V}}{\Omega R} Q^*_{\ell}(\Omega R) Y_{\ell m}(\theta,\phi).\label{sub-lv}
\end{eqnarray}
The functions $Q_{\ell}$ and $Q_{\ell}^*$ are already defined in \eqref{qell}. Once again, we see that these coefficients represent backscattering between the incoming and outgoing field modes in \eqref{fide}.


\subsection{Super-Hubble region}
We follow the same procedure as in the previous case, now with the metric \eqref{stsup} to obtain the following field equation for the super-Hubble region
\begin{equation}
    \Phi_\Omega = \sum_{\ell,m}\frac{\Phi^{\ell m}_\Omega(T,R)}{T}Y_{\ell m}(\theta, \phi)
\end{equation} 
since the radius of the two sphere is now given by $T$ and not $R$.  With the above expansion, the $R,T$ dependent part of the field equation is given by
\begin{equation}\label{fe2}
    \Bigg(\frac{\partial^2\Phi^{\ell m}_\Omega}{\partial T^2}- \frac{\partial^2\Phi^{\ell m}_\Omega}{\partial R^2}\Bigg) + \frac{\ell(\ell+1)}{T^2(H^2T^2-1)}\Phi^{\ell m}_\Omega = 0.
\end{equation}
Notice that the potential has a different form, as is expected because of the change in the metric coefficients in the super-Hubble region, given by
\begin{equation}
    V_{\ell} (T,H) = \frac{\ell(\ell+1)}{T^2(H^2T^2-1)}.
    \label{pot2}
\end{equation}
Once again, the potential diverges at the origin of the radial coordinate (which is simply $T$ for this region) and also at $T=1/H$. If we solve $T=1/H(T,R)$ we shall end up getting $R=1/H$ which is again telling us that there is a divergence at the Hubble radius. Consequently, once again, we see a peculiar behaviour just like the sub-Hubble case --  the $s=0$ modes do not see this potential but for $\ell>0$ all modes are confined outside the Hubble radius and cannot enter the sub-Hubble region.

\subsubsection{Field modes for the s-wave approximation}

For $\ell=0$, the solution of \eqref{fe2} is straightforward and the mode expansion is
\begin{equation}
    \Phi = \int d^3\Omega (U^{\text{sup}}_\Omega b^{in}_{\Omega<\Omega_H}+V^{\text{sup}}_\Omega b^{out}_{\Omega<\Omega_H}) + \text{h.c.}
\end{equation}
which is natural in the super-Hubble region. Particular expressions for the incoming and outgoing modes are
\begin{equation}\label{USupMode}
    U^{\text{sup}}_\Omega = \frac{1}{4\pi\sqrt{\Omega}T}e^{-i\Omega(T-R)}
\end{equation}
and
\begin{equation}\label{VSupMode}
    V^{\text{sup}}_\Omega = \frac{1}{4\pi\sqrt{\Omega}T}e^{-i\Omega(T+R)}.
\end{equation}
Once again the quantum state annihilated by $b^{in}_{\Omega<\Omega_H}$ and $b^{in}_{\Omega<\Omega_H}$ is the definition of $T-$vacuum in super-Hubble scale for respective in and out sectors.

\subsubsection{Field modes beyond the s-wave approximation}

Just like the sub-Hubble case the form of the potential \eqref{pot2} makes the differential equation hard to solve in general. But it is simpler if we are interested in deep super-Hubble modes for which $T<<1/H$, and \eqref{fe2}  takes the following form,
\begin{equation}\label{fe3}
    \Bigg(\frac{\partial^2\Phi^{\ell m}_\Omega}{\partial T^2}- \frac{\partial^2\Phi^{\ell m}_\Omega}{\partial R^2}\Bigg) - \frac{\ell(\ell+1)}{T^2}\Phi^{\ell m}_\Omega = 0.
\end{equation}
This is identical to \eqref{dsh} with an exchange $R\leftrightarrow T$, and therefore has the same solution \eqref{fide} -- we just need to replace $R$ with $T$. The final expression, with an undetermined normalization constant ${\cal N}_{\text{sup}}$, is now given by
\begin{widetext}
\begin{equation}
\label{fide2}
    \Phi = \sum_{\ell,m} \int d^3\Omega \,  {\cal N}_{\text{sup}} \sqrt{\frac{\Omega}{2\pi}} \left(b^{in}_{\Omega \ell m} \frac{e^{-i\Omega(T - R)}}{\Omega T} Q_{\ell}^*(\Omega T) + b^{out}_{\Omega \ell m} \frac{e^{-i\Omega(T + R)}}{\Omega T} Q_{\ell}(\Omega T) \right) Y_{\ell m}(\theta,\phi) + \text{h.c.}
\end{equation}
\end{widetext}
We can also extract the mode functions for the above expansion which are
\begin{eqnarray}\label{sup-lu}
    U_{\Omega \ell m} &=& \sqrt{\frac{\Omega}{4\pi}}  \frac{e^{-i\Omega U}}{\Omega T} Q_{\ell}^*(\Omega T) Y_{\ell m}(\theta,\phi) \\
    V_{\Omega \ell m} &=& \sqrt{\frac{\Omega}{4\pi}}  \frac{e^{-i\Omega V}}{\Omega T} Q_{\ell}(\Omega T) Y_{\ell m}(\theta,\phi).\label{sup-lv}
\end{eqnarray}

To conclude, we now have complete set of mode functions for the spherically symmetric form of the radiation dominated universe for the following cases -- (i) exact expressions \eqref{sub-su}, \eqref{sub-sv} and \eqref{USupMode}, \eqref{VSupMode} for the $s-$waves both in the sub and super-Hubble regions, (ii) approximate analytical solutions (up to undetermined normalization constants) \eqref{sub-lu}, \eqref{sub-lv} and \eqref{sup-lu}, \eqref{sup-lv} for the deep sub-Hubble and deep super-Hubble modes for $\ell>0$. We now move to the next section to use these results and calculate particle creation from $T-$vacuum.

\section{Particle creation from the $T-$vacuum}
\label{sec-5}

We are interested to calculate the number density of particles, created by  $T-$vacuum, when viewed from a cosmological frame of reference. The first step is to calculate the Bogolyubov coefficient $\beta_{\omega\Omega}$ and then use it to get the particle number density. 

By definition, the Bogolyubov coefficient has the following forms \cite{nsalas}
\begin{equation}
    \beta_{\omega\Omega}^{in} = -(u_\omega, U^*_\Omega) =i\int_{\Sigma}d\Sigma^\mu \: (u_\omega \partial_\mu U_\Omega - U_\Omega \partial_\mu u_{\omega}). 
\end{equation}
and
\begin{equation}
    \beta_{\omega\Omega}^{out} = -(v_\omega, V^*_\Omega) =i\int_{\Sigma}d\Sigma^\mu \: (v_\omega \partial_\mu V_\Omega - v_\Omega \partial_\mu v_{\omega}). 
\end{equation}
for the incoming and outgoing sectors. In the above expressions  $\Sigma$ is a Cauchy hypersurface and $d\Sigma^\mu = n^\mu d\Sigma$; $d\Sigma$ is the volume element associated with this Cauchy hypersurface and $n^\mu$ is a future directed unit normal vector to $\Sigma$. In this case, we are taking into account scalar field $\Phi$, which can be expanded into field modes as expressed in cosmological and $(T,R)$ frames in section \ref{sec-4}.
 
\subsection{Particle number density for the s-wave approximation}
Let us first consider the $s-$wave modes which are given by \eqref{uMode} and \eqref{vMode} for the cosmological frame, whereas, for the $(T,R)$ frame we have relevant expressions for the sub and super Hubble regions (appearing in \eqref{sub-su}, \eqref{sub-sv} and \eqref{USupMode}, \eqref{VSupMode}). In both coordinates modes are separated into incoming and outgoing sectors. 

We first select the incoming modes which are functions of retarded light-cone coordinates $u$ or $U$. Further, we consider the sub-Hubble modes for which we should choose \eqref{uMode} and \eqref{sub-su} for the calculation of Bogolyubov coefficient. It is useful to consider the  future null infinity  $\mathcal{I}^+$ as the Cauchy hypersurface.  This choice leads us to $d\Sigma=a^2(\eta)r^2dud\bar{\Omega}$ where $d\bar{\Omega}$ is the solid angle element. Therefore, 
\begin{equation}\label{BetaCoefficientIntegral}
    \beta_{\omega\Omega}^{in} = -(u_\omega, U^*_\Omega) = i\int_{\mathcal{I}^+} a^2(\eta)r^2dud\Omega \: (u_\omega\partial_uU_\Omega - U_\Omega\partial_u u_\omega).
\end{equation}
Doing an integration by parts and taking the limit corresponding the choice of the hypersurface ($\eta \rightarrow +\infty$ and $r \rightarrow +\infty$), we can write (\ref{BetaCoefficientIntegral}) as
\begin{equation}\label{BetaCoefficientIntegral2}
    \beta^{in}_{\omega\Omega} = 2i\int_{\mathcal{I}^+}a^2(\eta)r^2\: dud\bar{\Omega} \: u_{\omega}\partial_u(U_\Omega).
\end{equation}
with a scalar factor given by $a(\eta) = \mathcal{H}e\eta$. Simplifying (\ref{BetaCoefficientIntegral2}) and taking into account that $U = \pm \frac{\mathcal{H}e}{2}u^2$ for $u \geq 0$ and $u\leq 0$ respectively, we can express $\beta_{\omega\Omega}$ as
\begin{equation}
    \beta^{in}_{\omega\Omega} = I_1 + I_2
\end{equation}
where
\begin{equation}
    I_1 = \frac{\mathcal{H}e}{2\pi}\sqrt{\frac{\Omega}{\omega}}\int^\infty_0du\: u e^{-i(\omega u + \frac{\Omega\mathcal{H}e}{2}u^2)}.
\end{equation}
and
\begin{align}
    I_2 &= -\frac{\mathcal{H}e}{2\pi}\sqrt{\frac{\Omega}{\omega}}\int^0_{-\infty}du\: u e^{-i(\omega u - \frac{\Omega\mathcal{H}e}{2}u^2)}\\
    & = I^*_1.
\end{align}
Now using the identity 
\begin{equation}
    \int^\infty_{x_0} dx \ x^{s-1}e^{-bx} = e^{-s\log b}\Gamma[s, x_0] ; \ \ \text{Re} (b), \ \text{Re} (s) > 0
\end{equation}
where $b = -i + \epsilon$ and $\log b = \log(1) - i\pi/2 = -i\pi/2$,  we can integrate $I_1$ to to obtain
\begin{equation}
\begin{aligned}
    \beta^{in}_{\omega\Omega} &= \frac{1}{\pi\sqrt{\Omega\omega}}\sin\Big(\frac{\omega^2}{2\Omega\mathcal{H}e}\Big)\Gamma\Big[1;\frac{\omega^2}{2\Omega\mathcal{H}e}\Big]\\
    & -\frac{1}{2\pi\Omega}\sqrt{\frac{\omega}{\mathcal{H}e}}\Big[1+\sin\Big(\frac{\omega^2}{\Omega\mathcal{H}e}\Big)\Big]^{1/2}\Gamma\Big[\frac{1}{2};\frac{\omega^2}{2\Omega\mathcal{H}e}\Big].
\end{aligned}
\end{equation}
The  squared value is now given by 
\begin{widetext}
\begin{equation}
    \begin{aligned}
        |\beta_{\omega\Omega}^{in}|^2 &= \frac{1}{\pi^2\Omega\omega}\sin^2\Big(\frac{\omega^2}{2\Omega\mathcal{H}e}\Big)\Gamma^2\Big[1;\frac{\omega^2}{2\Omega\mathcal{H}e}\Big] \\ 
        &+ \frac{\omega}{4\pi^2\Omega^2\mathcal{H}e}\Big[1+\sin\Big(\frac{\omega^2}{\Omega\mathcal{H}e}\Big)\Big]\Gamma^2\Big[\frac{1}{2};\frac{\omega^2}{2\Omega\mathcal{H}e}\Big] \\
       & - \frac{1}{\pi^2\Omega^{3/2}\sqrt{\mathcal{H}e}}\sin\Big(\frac{\omega^2}{2\Omega\mathcal{H}e}\Big)\Big[1+\sin\Big(\frac{\omega^2}{\Omega\mathcal{H}e}\Big)\Big]^{1/2}\Gamma\Big[1;\frac{\omega^2}{2\Omega\mathcal{H}e}\Big]\Gamma\Big[\frac{1}{2};\frac{\omega^2}{2\Omega\mathcal{H}e}\Big].
    \end{aligned}
    \label{BogolyubovSquared4D}
\end{equation}
\end{widetext}
We can notice that the above result for $|\beta_{\omega\Omega}^{in}|^2$ is close (but not identical) to the  (1+1) dimensional problem which was calculated earlier in \cite{Modak:2018usa}, which is expected since radial modes in (3+1) dimensions is most close to a (1+1) dimensional problem.  The cross terms in \eqref{BogolyubovSquared4D} is a characteristic of the  (3+1) dimensional case. 

\begin{figure}[t]
    \centering
    \includegraphics[width = 0.6\linewidth]{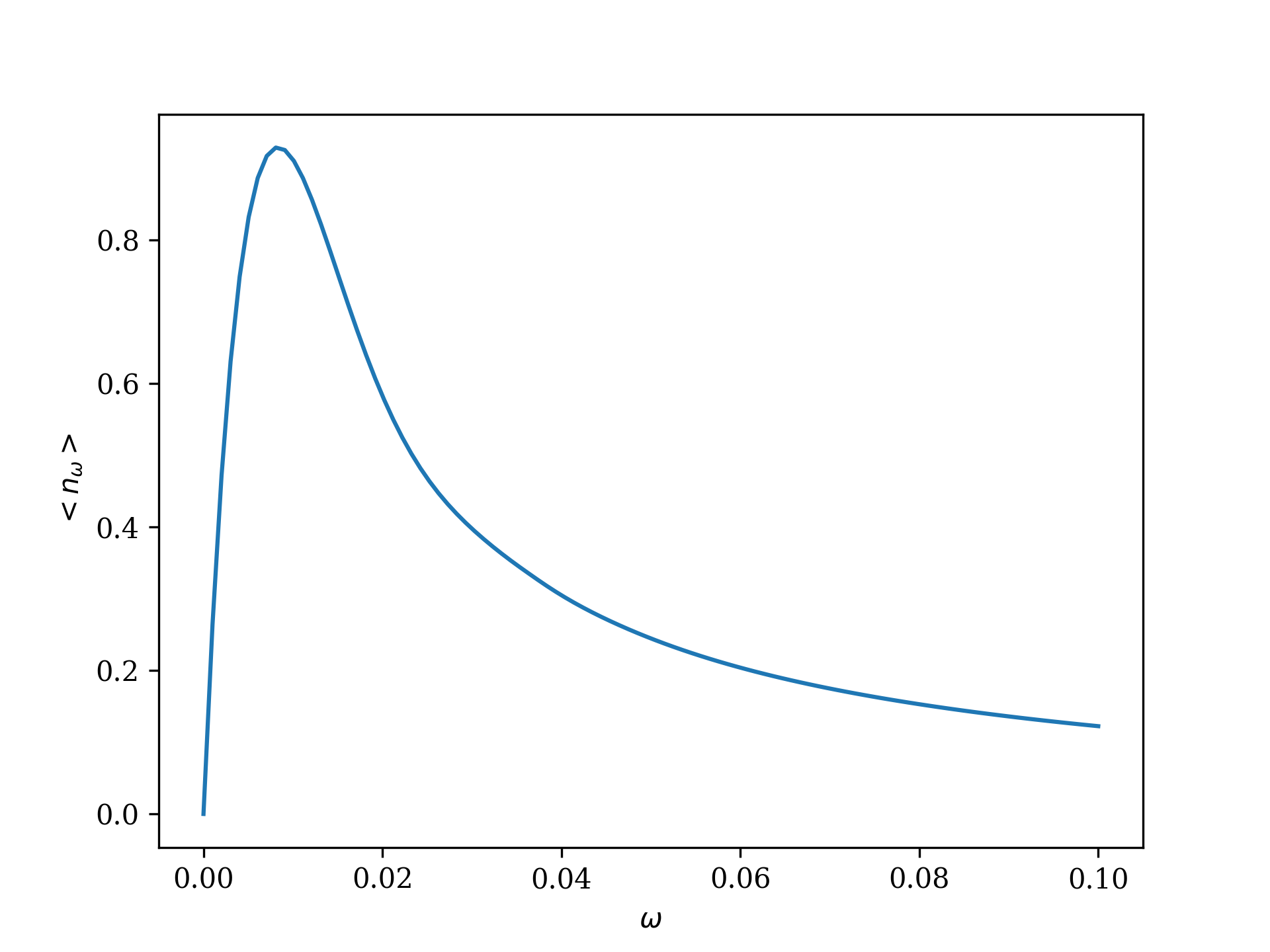}
    \caption{Plot of the average particle number density for modes fields in 3+1 dimensions versus the frequency $\omega$, from 0 to 0.1. This numerical solution is made using an infrared cut-off of $\Omega=0.0001$ and taking $\mathcal{H}=1$.}
    \label{NumberParticleDensity4D}
\end{figure}

The final step to get the particle number density is to integrate the last equation 
\begin{equation}
    \langle 0 | n_\omega |0 \rangle_{T} = \int |\beta_{\omega\Omega}^{in}|^2 d\Omega
\end{equation}
which unfortunately does not provide us a complete analytical solution. But it is easy to plot the average particle number density by performing  a numerical integration. It is evident from \eqref{BogolyubovSquared4D} that the particle number density has a infrared (IR) divergence which also appear in other cases of particle creation  such as in moving mirrors \cite{mir1, mir2}. We need to neglect modes with infinite wavelengths, and  an IR cut-off is used for performing the integration numerically. It is not a problem for our case since the universe will transit to the matter dominated stage after the radiation stage and the size of the universe at the transition will set a natural UV cut-off for the modes originated in the radiation stage. To perform the integration, here we put an ad-hoc value $\Omega=0.0001$ for IR cut-off. Further we put ${\cal H}=1$ for the plot which is shown  in Figure \ref{NumberParticleDensity4D}. We again see the behavior closely analogous to the (1+1) dimensional case reported previously \cite{Modak:2018usa}. It is easy to check that if one considers the outgoing modes \eqref{vMode} and \eqref{sub-sv} to calculate Bogolyubov coefficient $\beta_{\Omega \omega}^{out}$ and then use it to calculate the particle content for outgoing sector in sub-Hubble region, we once again obtain a result identical to the incoming sector and the plot \ref{NumberParticleDensity4D} still applies.

Notice that this result is obtained from the contribution of sub-Hubble field modes $U^{\text{sub}}_\Omega$, and we can also look for the result in the super-Hubble sector. If we now do the same calculation for the super-Hubble modes $U^{\text{sup}}_\Omega$ in \eqref{USupMode} and \eqref{uMode}, together with the corresponding transformation \eqref{TransformationRegion2} we can realize that we obtain the same integral for the relevant Bogolyubov coefficient, and therefore, we shall obtain the same result, as depicted in Fig. \ref{NumberParticleDensity4D} for the incoming sector in super-Hubble region. The same is true for the outgoing $v-$modes in the super-Hubble region.  Therefore, we conclude that the plot \ref{NumberParticleDensity4D} is generic and applies for both the sub and super Hubble regions, and for the incoming and the outgoing sectors.

\subsection{Particle number density beyond the s-wave approximation}

Let us now compute the particle number density for the most general case with $\ell >0$. For this case we don't have the luxury to calculate the Bogolyubov coefficients or the particle content with all generalities. The only situations where mode decomposition is possible in the $(T,R)$ frame are away from the Hubble radius, that is, for both the deep sub-Hubble and the deep super-Hubble modes ($R<<1/H$ and $R>>1/H$). These field expansions are shown in \eqref{fide} and \eqref{fide2}. On the other hand, in the cosmological coordinates, mode decomposition is valid for all sub-Hubble and super-Hubble scale as appear in \eqref{fideo}. Taking into account first the incoming sector ($u-$ modes) and the deep sub-Hubble region, we can express the Bogolyubov coefficient as \footnote{We shall use the unnormalized modes for the spherically symmetric FRW by dropping ${\cal N}_{\text{sub/sup}}$ in \eqref{fide} and \eqref{fide2} whose exact determination needs a numerical calculation. We note that normalization will only change the scale of the final result and not on the form or relative intensity among various modes.}
\begin{widetext}
\begin{equation}\label{bog2}
   \beta^{in}_{\omega\Omega\ell \ell' m m'} = i\int a^2 r^2 dr \sin\theta d\theta d\phi (u_{\omega\ell m} \partial_\eta U_{\Omega \ell' m'} - U_{\Omega\ell' m'} \partial_\eta u_{\omega\ell m}).
\end{equation}
\end{widetext}
An explicit calculation of the integral \eqref{bog2} is lengthy and it is performed in the appendix \ref{ap-b}. It is nice to see a closed form analytical result which is given by 
\begin{widetext}
\begin{equation}\label{bogcof}
    |\beta^{in}_{\omega\Omega\ell}|^2  = \frac{\eta \omega^2}{4\mathcal{H}e\Omega^2}J^2_{\ell-1/2}(\eta \omega) - \frac{2\ell \omega}{4\mathcal{H}e\Omega^2}J_{\ell-1/2}(\eta \omega)J_{\ell+1/2}(\eta \omega)+ \frac{\ell^2+\eta^2\omega^2}{4\mathcal{H}e \Omega^2 \eta} J^2_{\ell+1/2}(\eta \omega).
\end{equation}
\end{widetext}
The average number density of particles created by the $T-$vacuum is therefore given by 
\begin{equation}\label{noop}
    n_{\omega\ell} = \int{ |\beta^{in}_{\omega\Omega\ell}|^2  d\Omega}.
\end{equation}
Considering \eqref{bogcof} it is straightforward to integrate and obtain once again a close form expression for the particle content, given by
\begin{widetext}
\begin{equation}\label{noop2}
\begin{aligned}
    \langle 0| n_{\omega\ell} |0\rangle_T  &= \frac{1}{4{\cal H} e \Omega_0 \eta} \left(\eta^2 \omega^2 J^2_{\ell-1/2}(\eta \omega) - {2\ell \omega \eta}J_{\ell-1/2}(\eta \omega)J_{\ell+1/2}(\eta \omega) \right.\\
     & \left. + (\ell^2 + \eta^2\omega^2) {J^2_{\ell+1/2}(\eta \omega)} \right),
\end{aligned}
\end{equation}
\end{widetext}
where $\Omega_0$ is again the IR cutoff introduced in the integration limit in equation \eqref{noop}. As we discussed in the 1+1 dimensional situation that the IR cut-off naturally applies here since the size of the universe in the radiation stage remain when it transits to the matter dominated stage just before producing the CMB radiation. We can therefore safely ignore modes of infinite wavelengths. Note that there is no ultra-violet divergence here which is reassuring. Finally, we want to use the scale factor as the time parameter  by using  its relationship with conformal time parameter given by $a(\eta) = {\cal H} e \eta$ for the radiation stage, and then the expression  \eqref{noop2} takes the following form
\begin{widetext}
\begin{equation}\label{noop3}
\begin{aligned}
    \langle 0|_T n_{\omega\ell} |0\rangle_T  &= \frac{1}{4 a \Omega_0} \left( (a \omega/{\cal H}e)^2 J^2_{\ell-1/2}(a \omega/{\cal H}e) - \frac{2\ell a \omega}{{\cal H} e} J_{\ell-1/2}(a \omega/{\cal H}e)J_{\ell+1/2}(a \omega/{\cal H}e) \right.\\
     & \left. + (\ell^2 + a^2\omega^2/{\cal H}^2e^2) {J^2_{\ell+1/2}(a \omega/{\cal H}e)} \right).
\end{aligned}
\end{equation}
\end{widetext}

\begin{figure}[t]
    \centering
    \includegraphics[width = 1\linewidth]{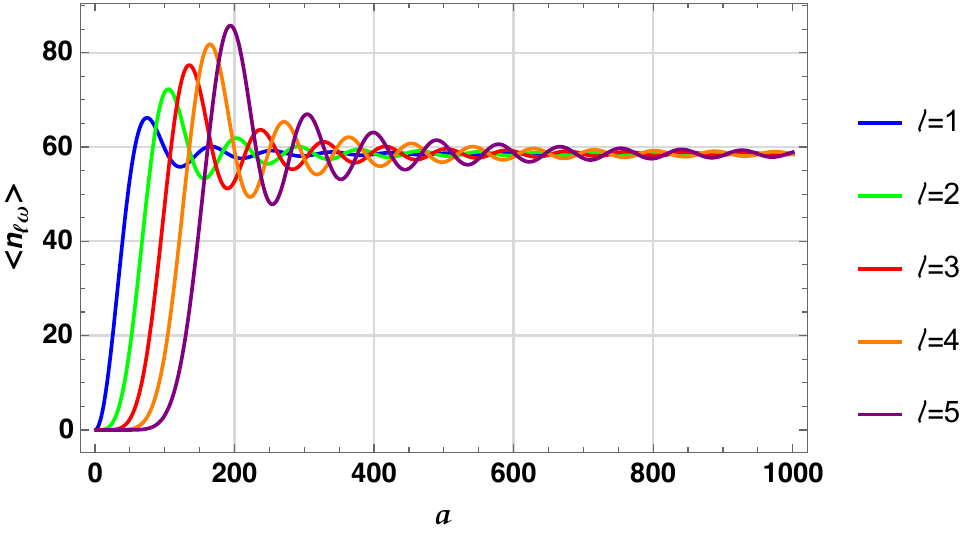}
    \caption{Plot of the particle number density with the scale factor as appear in equation \eqref{noop3}. Particle number oscillates with the scale factor during the expansion of the universe with a diminishing amplitude of oscillation and finally approaching a saturation towards the end of the radiation epoch.}
    \label{nlo1}
\end{figure}

A few comments are in order regarding our final result \eqref{noop3} -- first and foremost, the particle content is manifestly anisotropic and now known for any value of the angular momentum quantum number $\ell>0$. This is a direct consequence of the fact that the $T-$vacuum is intrinsically anisotropic as it is a natural state for observers who envision the radiation dominated universe as such. Of course, these observers, who are the successors of static de Sitter observers, do not see any particles as the quantum state is in vacuum configuration for them. However, the cosmological observers with proper time $\eta$ are bombarded with particles due its peculiar motion with respect to $(T,R)$ frame. These particles are anisotropically distributed in the sky and its distribution is given by \eqref{noop3}. Evidently, this is not a blackbody distribution and it is not at all a surprise since the relationship between the cosmological and $(T,R)$ frames do not invoke an exponential redshift relative to each other -- rather it is a power law and quadratic in nature. Furthermore, \eqref{noop3} is time dependent and it has an oscillatory behavior with the scale factor which we depict in Figure \eqref{nlo1}.

Notice again that if we consider the outgoing $v-$modes in \eqref{vMode} and \eqref{sub-lv} we shall end up with the same result as \eqref{noop3}. In addition, the result for the deep super-Hubble modes \eqref{sup-lu} and \eqref{sup-lv} lead us to an identical result \eqref{noop3}. Thus we can conclude that \eqref{noop3} and figure \eqref{nlo1} are generic behavior of the anisotropic particle creation by the $T-$vacuum in appropriate limits.

\section{Discussions and Outlook}
\label{sec-6}

We made a significant step forward making a case in favor of a physically appealing, anisotropic quantum vacuum state in the radiation dominated stage of the early universe. To our knowledge this is the only quantum vacuum state whose excitation most naturally breaks the homogeneity and isotropy for cosmological observers in the post inflationary era. Our results are however valid in the sub/super Hubble regions away from the Hubble radius which is interesting enough given our own location in the cosmos. We considered massless scalar fields for this study and quantized them in the two different reference frames - (i) standard cosmological frame, (ii) a recently discovered spherically symmetric form of FRW defining the radiation stage. The latter frame was identified, by back-tracking, with the static de Sitter frame in the inflationary stage. Quantum fields, while in curved spacetime, are allowed to have non-unique vacuum states, although they might occupy only one of such physically allowed states while remain unexcited. The presence of $T-$vacuum and the very excitation of it provides an observational angular variance of radiation to an cosmological observer and such observers are closely related to the comoving observers whose frame shapes our own understanding of cosmological observations.

This particular nature of $T-$vacuum raises a few interesting theoretical implications and it is worth investigating further on it. One natural ambition might be calculating some definite imprints of these anisotropies originating from the $T-$vacuum and compare them with the CMB map. This prospect is borne by the fact that the CMB map depicts the early universe in the matter dominated stage which is after the radiation stage and therefore a suitable signature in the radiation stage, such as in \eqref{noop3}, may have some residual imprint on the CMB map itself. It provides exact number of particles (such as photons) corresponding to a given frequency and direction in the sky. This prediction is extremely precise and not to be confused with the statistical average, which is usually done while analysing CMB data, in the lack of a precise physical theory that might be producing the CMB radiation. Of course, we do not claim anything about the observed CMB anisotropy with the anisotropy calculated in \eqref{noop3} yet, but it is curious that such an expression can be calculated from such a first principle theory where a anisotropic photon background might be related with the symmetry of quantum vacuum itself. It would be interesting to see if the calculated anisotropies generated by $T$-vacuum can eventually be compared with CMB anisotropies, and in addition if these new particles could play some part in reheating the early universe. We are currently investigating these aspects within a numerical set up and  plan to report the outcomes in a future publication \cite{ams}.

\section{Acknowledgments}
One of the authors (SKM) thanks Facultad de Ciencias, UCOL and ICN-UNAM for providing research spaces for this work. ES thanks CONAHCyT for financial support and SKM acknowledges start-up Fund from Cal Poly Pomona.

\appendix
\section{The normalization constant for cosmological modes}\label{ap-a}
Since the solutions for the cosmological mode functions derived in Section \ref{sec-3b} are exact, we can easily normalize them by applying the inner product to the field modes:
\begin{align}\label{norm}
    (u_{\omega1}, u_{\omega2})  &=  -i  \mathcal{N}^2 \int d\Sigma \frac{\sqrt{-g{_{_\Sigma}}}}{a}\sum_{\ell,m,\ell',m'} [\frac{e^{-i\omega_1\eta}}{\omega_1 \eta} \frac{J_{\nu}(\omega_1 r)}{\sqrt{r}} Y_{\ell m}(\theta,\phi) (i\omega_2 -\frac{1}{\eta}) \frac{e^{i\omega_2\eta}}{\omega_2 \eta} \frac{J_{\nu'}(\omega_2 r)}{\sqrt{r}}\nonumber\\
    & \times Y^*_{\ell'm'}(\theta,\phi) - \frac{e^{i\omega_2\eta}}{\omega_2\eta} \frac{J_{\nu'}(\omega_2 r)}{\sqrt{r}} Y^*_{\ell'm'}(\theta,\phi) (-i\omega_1 -\frac{1}{\eta}) \frac{e^{-i\omega_1\eta}}{\omega_1 \eta} \frac{J_{\nu}(\omega_1 r)}{\sqrt{r}} Y_{\ell m}(\theta,\phi) ] \nonumber \\
    & =  \mathcal{N}^2 \int \sum_{\ell,m,\ell',m'}(\mathcal{H}e\eta )^2r^2\sin\theta dr d\theta d\phi \frac{e^{-i(\omega_1-\omega_2)\eta}}{\omega_1 \omega_2 \eta^2} \nonumber \\ 
    & \qquad \times \frac{J_{\nu}(\omega_1 r) J_{\nu'}(\omega_2 r)}{r} Y_{\ell m}(\theta,\phi) Y^*_{\ell'm'}(\theta,\phi)(\omega_1+\omega_2) \nonumber \\
    & = \mathcal{N}^2 \int \sum_{\ell,m,\ell',m'}(\mathcal{H}e)^2r^2 dr e^{-i(\omega_1-\omega_2)\eta} \frac{J_{\nu}(\omega_1 r) J_{\nu'}(\omega_2 r)}{r} \frac{(\omega_1+\omega_2)}{\omega_1 \omega_2}  \delta_{\ell \ell'}\delta_{mm'}\nonumber \\
    & = \mathcal{N}^2 \mathcal{H}^2 e^2 e^{-i(\omega_1-\omega_2)\eta}\frac{(\omega_1+\omega_2)}{\omega_1 \omega_2}  \int r dr J_{\nu}(\omega_1 r) J_{\nu}(\omega_2 r) \nonumber \\
    & =  \mathcal{N}^2 \mathcal{H}^2 e^2 e^{-i(\omega_1-\omega_2)\eta}\frac{(\omega_1+\omega_2)}{\omega_1 \omega_2} \times \frac{1}{\omega_1}\delta(\omega_1 -\omega_2 ) \nonumber \\
    \implies \mathcal{N} &  =  \frac{\omega }{\sqrt{2} \mathcal{H}e}.  
\end{align}

\section{Bogolyubov coefficient for $\ell > 0$}\label{ap-b}

The calculation of the Bogolyubov coefficient is performed by taking a space-like hypersurface at $\eta = \text{constant}$. It should be recalled that the scale factor is related to the conformal time by a linear relationship $a(\eta) = {\cal H} e\eta$. We start from the basic definition \eqref{bog2} 
%
\begin{align*}
    \beta_{\omega\Omega\ell \ell' m m'} & = i\int a^2 r^2 dr \sin\theta d\theta d\phi (u_{\omega\ell m} \partial_\eta U_{\Omega \ell' m'} - U_{\Omega\ell' m'} \partial_\eta u_{\omega\ell m})\\
    & = i \frac{\sqrt{\mathcal{H}e \eta}}{2} e^{-i(\omega \eta + \tfrac{1}{2}\mathcal{H}e \Omega\eta^2)} \int^\infty_0 dr\; r e^{-i\mathcal{H}e\Omega r^2/2} J_{\ell +1/2}(\omega r) \\
     & \times \left[2\eta \frac{d}{d\eta}J_{\ell' + 1/2}(\mathcal{H}e\Omega \eta r) + (1+2i\eta(\omega-\mathcal{H}e\Omega\eta))J_{\ell' + 1/2}(\mathcal{H}e\Omega \eta r) \right] \nonumber\\
    & \times \int_{\theta=0}^{\pi}\int_{\phi=0}^{2\pi} d\theta d\phi \sin\theta Y_{\ell m} (\theta,\phi) Y_{\ell' m'}(\theta,\phi)\\
 \implies \beta_{\omega\Omega\ell m}   &=i(-1)^m \frac{\sqrt{\mathcal{H}e \eta}}{4} e^{-i(\omega \eta + \tfrac{1}{2}\mathcal{H}e \Omega\eta^2)} \left[ 2\eta\frac{d}{d\eta} + (1+i2\eta(\omega-\mathcal{H}e\Omega\eta)) \right] \nonumber \\
 & \times \int^\infty_0 dr\; r e^{-i\mathcal{H}e\Omega r^2/2} J_{\ell + 1/2}(\omega r) J_{\ell + 1/2}(\mathcal{H}e\Omega \eta r) \nonumber \\
    &= i(-1)^m \frac{\sqrt{\mathcal{H}e \eta}}{2} e^{-i(\omega \eta + \tfrac{1}{2}\mathcal{H}e \Omega\eta^2)}\left[ 2\eta\frac{d}{d\eta} + (1+i2\eta(\omega-\mathcal{H}e\Omega\eta)) \right] \\
    & \times \frac{-i}{\mathcal{H}e\Omega} e^{i \frac{\omega^2+\mathcal{H}^2e^2\Omega^2\eta^2}{2\mathcal{H}e\Omega}} I_{\ell+1/2}\left(-i\omega \eta \right)\\
    & = (-1)^m\frac{\sqrt{\mathcal{H}e \eta}}{2\mathcal{H}e \Omega}e^{-i\omega\eta+i\frac{(\omega^2+\mathcal{H}^2e^2\Omega^2\eta^2)}{2\mathcal{H}e\Omega } } \left[ 2\eta\frac{d}{d\eta} + (1+i2\eta(\omega-\mathcal{H}e\Omega\eta)) \right]  I_{\ell+1/2}(-i \omega\eta),
\end{align*}
where $I_{\ell+1/2}(\eta\omega)$ is modified Bessel function of first kind, such that $I_{\ell+1/2}(z) = e^{-i(\ell+1/2)\pi/2} J_{\ell+1/2}(iz)$. From the above expression we obtain \eqref{bogcof}.

\end{document}